\documentclass[preprint,amsmath,amssymb,aps]{revtex4-1}
\usepackage{graphicx}
\usepackage{dcolumn}
\usepackage{bm}

\begin{document}

\title{Raman Spectrum of CrI$_3$: an \emph{ab-initio} study}

\author{Daniel Larson} \affiliation{Department of Physics, Harvard
  University, Cambridge, Massachusetts, 02138, USA.}
\author{Efthimios Kaxiras} \affiliation{Department of Physics, Harvard
  University, Cambridge, Massachusetts, 02138, USA.}
\affiliation{John A. Paulson School of Engineering and Applied
  Sciences, Harvard University, Cambridge, Massachusetts, 02138, USA.}

\date{\today}

\begin{abstract}
  We study the Raman spectrum of CrI$_3$, a material that exhibits
  magnetism in a single-layer. We employ first-principles
  calculations within density functional theory to determine the
  effects of polarization, strain, and incident angle on the phonon
  spectra of the 3D bulk and the single-layer 2D structure, for both
  the high- and low-temperature crystal structures. Our results are in
  good agreement with existing experimental measurements and
  serve as a guide for additional investigations to elucidate the
  physics of this interesting material.
\end{abstract}

\maketitle

The demonstration of magnetism in a single layer of
CrI$_3$~\cite{huang2017layer} has exciting implications for the
creation of novel materials and devices based on combinations of
atomic layers with varied properties. CrI$_3$ is a layered material
that can be cleaved into flakes varying in thickness from a single
layer to many layers. A standard, powerful technique to characterize
the properties of atomically thin materials is Raman
spectroscopy~\cite{zhang2016review}. Motivated by the importance of Raman
spectroscopy experiments, we have studied the phonon spectrum of bulk
and single-layer CrI$_3$ in order to elucidate the dependence of the
Raman spectrum on thickness, polarization, strain, and incident
angle. Our results agree well with the location and polarization
dependence of the Raman peaks in existing experimental measurements
and serve as a guide for further studies of this intriguing material.

We obtain the Raman spectrum of CrI$_3$ using first-principles density
functional theory (DFT) calculations. In order to study the dependence
of the Raman spectrum on the sample thickness, we considered periodic
cells with crystallographic dimensions and cells containing a single
layer separated from its periodic images in the direction
perpendicular to the layer by a vacuum region of 20 \AA, which
eliminates interlayer interactions.

Starting from the crystallographic structure, we obtain the fully
relaxed primitive cell parameters using
VASP~\cite{Kresse1996,Kresse1996a} spin-polarized calculations with
Projector Augmented Wave (PAW) pseudo-potentials, the PBE
exchange-correlation functional~\cite{Perdew1996}, a plane-wave energy
cutoff of 300 eV, and a 10$\times$10$\times$10 $k$-point mesh
(10$\times$10$\times$1 for single-layer structures). Van der Waals
corrections were taken into account using the DFT-D3
method~\cite{Grimme2010}.

We determine the phonon frequencies and normal modes at the $\Gamma$ point from
the internal forces of 2$\times$2$\times$2 (2$\times$2$\times$1) bulk
(single-layer) supercells containing displaced atoms, using the phonopy
code~\cite{Togo2015}, with a properly scaled $k$-point mesh.
To estimate the relative intensity of Raman scattering from each
mode we calculate the derivative of the 
macroscopic dielectric tensor based on finite differences for cells
displaced in the direction of each mode~\cite{Fonari2013}.

\begin{figure}
  \includegraphics[width=\columnwidth]{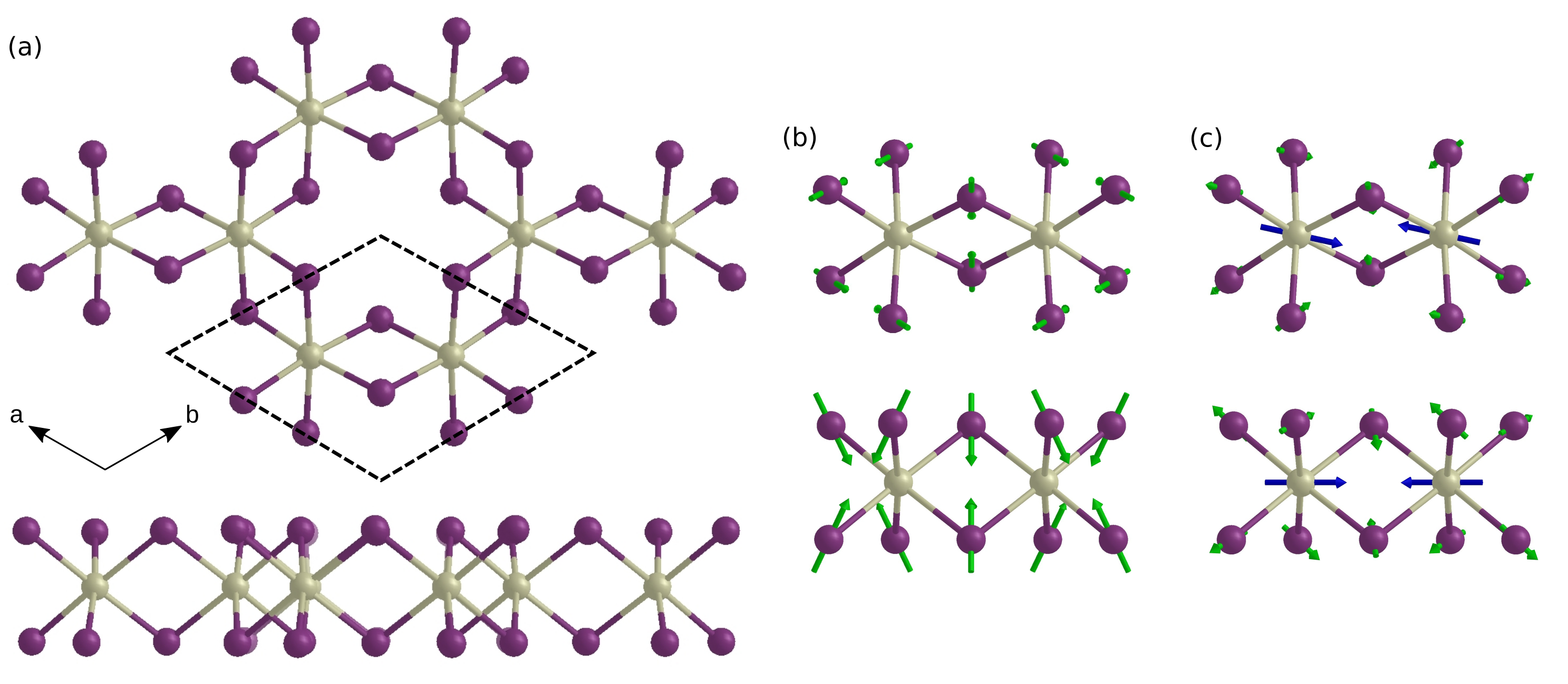}
\caption{(Color online) (a) Top and side views of the crystal structure of
  single-layer CrI$_3$. Iodine atoms are purple and chromium atoms are
  grey. The dashed line shows a 2D primitive unit cell. (b) Top and
  side views of the $A_{1g}^2$ mode (110 cm$^{-1}$) atomic
  displacements for a single layer of the R$\overline{3}$
  structure. Displacements are proportional to the length of the arrow
  (blue for Cr, green for I). This mode corresponds to an alternate
  squeezing and spreading of the I layers in the $z$-direction with
  the Cr layer fixed. (c) $E_g^4$ (223 cm$^{-1}$) mode atomic
  displacements are dominated by alternate in-plane motion of the Cr
  ions. For the other member of the $E_g^4$ doublet the Cr ions move
  in the perpendicular direction within the plane of the layer.}
\label{fig:struct}
\end{figure}

At room temperature CrI$_3$ has the monoclinic AlCl$_3$ structure type
(space group C2/m, \#12), but as the temperature decreases below
$T\sim 220$ K it undergoes a structural phase transition to the
rhombohedral BiI$_3$ structure type (space group R$\overline{3}$,
\#148)~\cite{McGuire2015}. The crystal structure of single-layer CrI$_3$
is shown in Fig.~\ref{fig:struct}. The bulk structure consists
of stacks of the individual layers, with the primary difference between high-
(C2/m) and low-temperature (R$\overline{3}$) structures being the alignment of the
layers.

The phonon modes of CrI$_3$ are determined by the symmetries of the
point groups associated with the high- and low-temperature structures.
The irreducible representation of the point group to which each phonon
mode belongs determines whether or not it is Raman active. For the
high-temperature (C2/m) structure the point group is $C_{2h}$ which
has four one-dimensional irreducible representations, labeled $A_g$,
$B_g$, $A_u$, $B_u$. The $A_g$ and $B_g$ representations have even parity
under inversion, while the $A_g$ and $A_u$ modes are even under the $C_2$
rotation. The 24 modes of the phonon spectrum with $C_{2h}$ symmetry
are divided between the irreducible representations as follows:
\begin{eqnarray}
  \Gamma_{\mathrm{acoustic}} &=& A_u + 2 B_u \\
  \Gamma_\mathrm{optical} &=& 6 A_g + 6 B_g + 4 A_u + 5 B_u.
\end{eqnarray}
For the low temperature (R$\overline{3}$) structure the point group is
$C_{3i}$ = S$_6$, which has two one-dimensional irreducible
representations $A_g$ and $A_u$ and two 2-dimensional irreducible
representations $E_g$ and $E_u$, each describing a pair of degenerate
modes. The phonon spectrum with $C_{3i}$ symmetry has the following
representations:
\begin{eqnarray}
\Gamma_{\mathrm{acoustic}} &=& A_u + E_u \\
\Gamma_\mathrm{optical} &=& 4 A_g + 4 E_g + 3 A_u + 3 E_u. 
\end{eqnarray}
As the sample thickness decreases, the differences between the phonon
spectra of the two structures will decrease. In fact, the single-layer
of both structures has a higher point symmetry than the bulk, namely
$D_{3d}$, most likely corresponding to the spacegroup R$\bar{3}2/m$
(\#166). The $D_{3d}$ point group has the following irreducible
representations:
\begin{eqnarray}
\Gamma_{\mathrm{acoustic}} &=& A_{2u} + E_u \\
\Gamma_\mathrm{optical} &=& 2 A_{1g} + 2 A_{2g} + 4 E_g +  A_{1u} + 2
A_{2u}+ 3 E_u.
\end{eqnarray}

The symmetries of the Raman tensor can be used to predict which modes
will be active and with what polarization. The typical experimental
setup is a back-scattering geometry where the incident light can be
plane polarized parallel to the surface of the sample and the
scattered light is measured through another polarizer that is either
parallel or perpendicular to the incident polarization. We consider
measurements with the incident light normal to the CrI$_3$ layers. All
Raman active modes must be inversion symmetric (``g'' modes). The
results of the first-principles calculations of the phonon spectrum of
CrI$_3$, including the polarization dependence, are shown in
Table~\ref{tab:modes}. As expected, the results from the two
calculations of the single-layer spectrum are quite similar. All
frequencies of the single-layer R$\overline{3}$ structure are slightly
lower than those of the corresponding modes in the single-layer C2/m
structure, but none by more than 2\%.

The bulk C2/m spectrum displays near degeneracy between several pairs
of $A$ and $B$ modes. By comparing to the single-layer and
R$\overline{3}$ spectra, we see that they match the $E$ modes of the
$C_{3i}$ and $D_{3d}$ point groups. This implies that the symmetry
breaking in the C2/m structure due to the layer stacking is not very
strong, as one might expect for weak van der Waals coupling between
the layers.

\begin{figure}
  \includegraphics[width=.8\columnwidth]{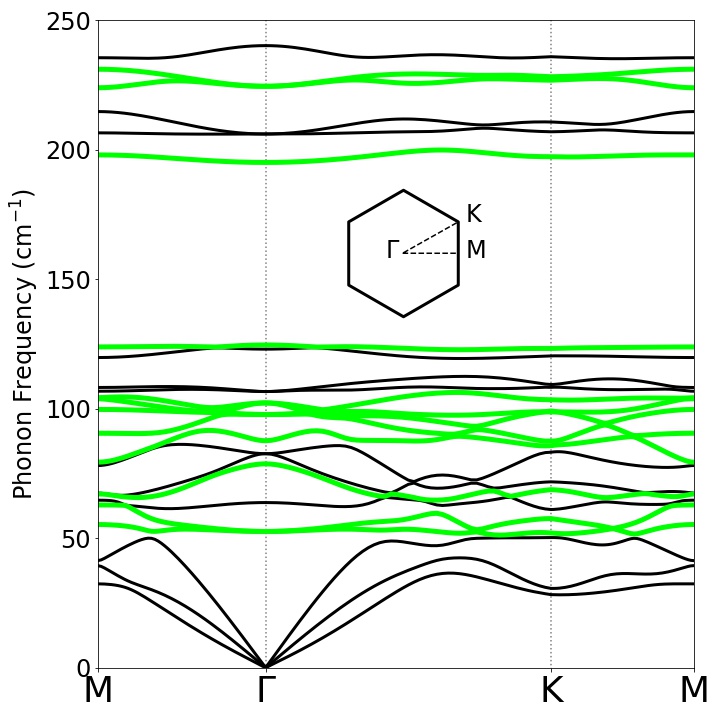}
\caption{(Color online) Phonon band structure for the bulk
  R$\overline{3}$ structure along a path from M to $\Gamma$ to K to M
  in the 2D Brillouin zone, shown in the inset. Colored bands show
  those modes expected to be Raman active at $\Gamma$.}
\label{fig:band}
\end{figure}

In Fig.~\ref{fig:struct}(b) and (c) we show the atomic displacements
for the two modes with largest measured intensity. In
Fig.~\ref{fig:band} we show the phonon frequencies along high symmetry
directions of the the Brillouin zone for the bulk R$\overline{3}$
structure. Modes that are Raman active at the $\Gamma$ point are
indicated in color.  The phonon modes for the C2/m structure are very
similar (almost indistinguishable on the scale of this plot). It is
interesting to note that the Raman active modes for the single-layer
structure are shifted to lower frequency as compared to the
corresponding bulk modes.

\begin{table*}[!ht]
  \centering
  \begin{tabular}{|c|r|r|c|r|r|c|} \hline
  \multicolumn{3}{|c|}{R$\overline{3}$ Structure} & & \multicolumn{3}{c|}{C2/m
    Structure} \\ \hline
  $C_{3i}$ & Bulk & Single-layer & $D_{3d}$ & Single-layer & Bulk & $C_{2h}$ \\
   & $\omega_\mathrm{opt}$ & $\omega_\mathrm{opt}$ &  &
  $\omega_\mathrm{opt}$ & $\omega_\mathrm{opt}$ &  \\ \hline\hline 
  $A_u$ & 240 & 245 & $A_{2u}$ & 246 & 242 & $B_u$ \\ 
  $E_g$ & $\parallel,\perp$ \textbf{225} & $\parallel,\perp$ $\mathbf{223}$ &
  $E_g$ & $\parallel, \perp$ \textbf{225} & 225 & $B_g$  \\
        & $\parallel,\perp \mathbf{225}$ & $\parallel,\perp
  \mathbf{223}$ & & $\parallel,\perp \mathbf{225}$ & $\parallel
  \mathbf{224}$ &  $A_g$ \\ 
  $E_u$ & 206 & 205 & $E_u$ & 207 & 206 &   $B_u$ \\
        & 206 & 205 &      & 207 & 206 &  $A_u$ \\
  $A_g$ & $\parallel \mathbf{195}$ & 195 & $A_{2g}$ & 196 & 196 &   $B_g$ \\ 
  $A_g$ & $\parallel \mathbf{125}$ & $\parallel \mathbf{119}$ &
  $A_{1g}$ & $\parallel \mathbf{120}$ & $\parallel \mathbf{125}$ &   $A_g$ \\
  $A_u$ & 123 & 122 & $A_{1u}$ & 123 & 124 &   $A_u$ \\
  $E_u$ & 107 & 105 & $E_u$ & 106 & 108 & $A_u$ \\
        & 107 & 105 &      & 106 & 107 & $B_u$  \\ 
  $E_g$ & $\parallel,\perp \mathbf{102}$ & $\parallel,\perp
  \mathbf{99}$ & $E_g$ & $\parallel,\perp \mathbf{100}$ & $\parallel
  \mathbf{102}$ &   $A_g$ \\  
  & $\parallel,\perp \mathbf{102}$ & $\parallel,\perp \mathbf{99}$ &
  &  $\parallel,\perp \mathbf{100}$ & 101 &   $B_g$ \\  
  $E_g$ & $\parallel,\perp \mathbf{98}$ & $\parallel,\perp
  \mathbf{96}$ & $E_g$ & $\parallel,\perp \mathbf{96}$ & $\parallel
  \mathbf{99}$ &   $A_g$  \\  
  & $\parallel,\perp \mathbf{98}$ & $\parallel,\perp \mathbf{96}$ &  &
  $\parallel,\perp \mathbf{96}$ & 99 & $B_g$  \\ 
  $A_g$ & $\parallel \mathbf{88}$ & 86 & $A_{2g}$ & 87 & $\perp
  \mathbf{86}$ & $B_g$  \\ 
  $E_u$ & 83 & 76 & $E_u$ & 77 & 82 &  $A_u$ \\ 
        & 83 & 76 &     &  77 & 79 &   $B_u$  \\ 
  $A_g$ & $\parallel \mathbf{79}$ & $\parallel \mathbf{75}$ & $A_{1g}$ &
  $\parallel \mathbf{76}$ & $\parallel \mathbf{79}$ &  $A_g$  \\
  $A_u$ & 64 & 58 & $A_{2u}$ & 59 & 58 &  $B_u$ \\ 
  $E_g$ & $\parallel,\perp \mathbf{53}$ & $\parallel,\perp
  \mathbf{48}$ & $E_g$ & $\parallel,\perp \mathbf{48}$ & 52 &  $B_g$  \\  
  & $\parallel,\perp \mathbf{53}$ & $\parallel,\perp \mathbf{48}$ &  &
  $\parallel,\perp \mathbf{48}$ & $\parallel \mathbf{51}$ &   $A_g$  \\ \hline 
\end{tabular}
  \caption{Calculated optical phonon frequencies
    $\omega_\mathrm{opt}$, in cm$^{-1}$, and assignment to irreducible
  representations of $C_{3i}$, the point group for the R$\overline{3}$
structure, $C_{2h}$, the point group of the C2/m structure, and
$D_{3d}$, the point group of the single-layer structure. Bold values
correspond to Raman active modes for parallel ($\parallel$) and
perpendicular ($\perp$) configurations of incident and scattered radiation.}
\label{tab:modes}
\end{table*}

\begin{figure}
  \includegraphics[width=\columnwidth]{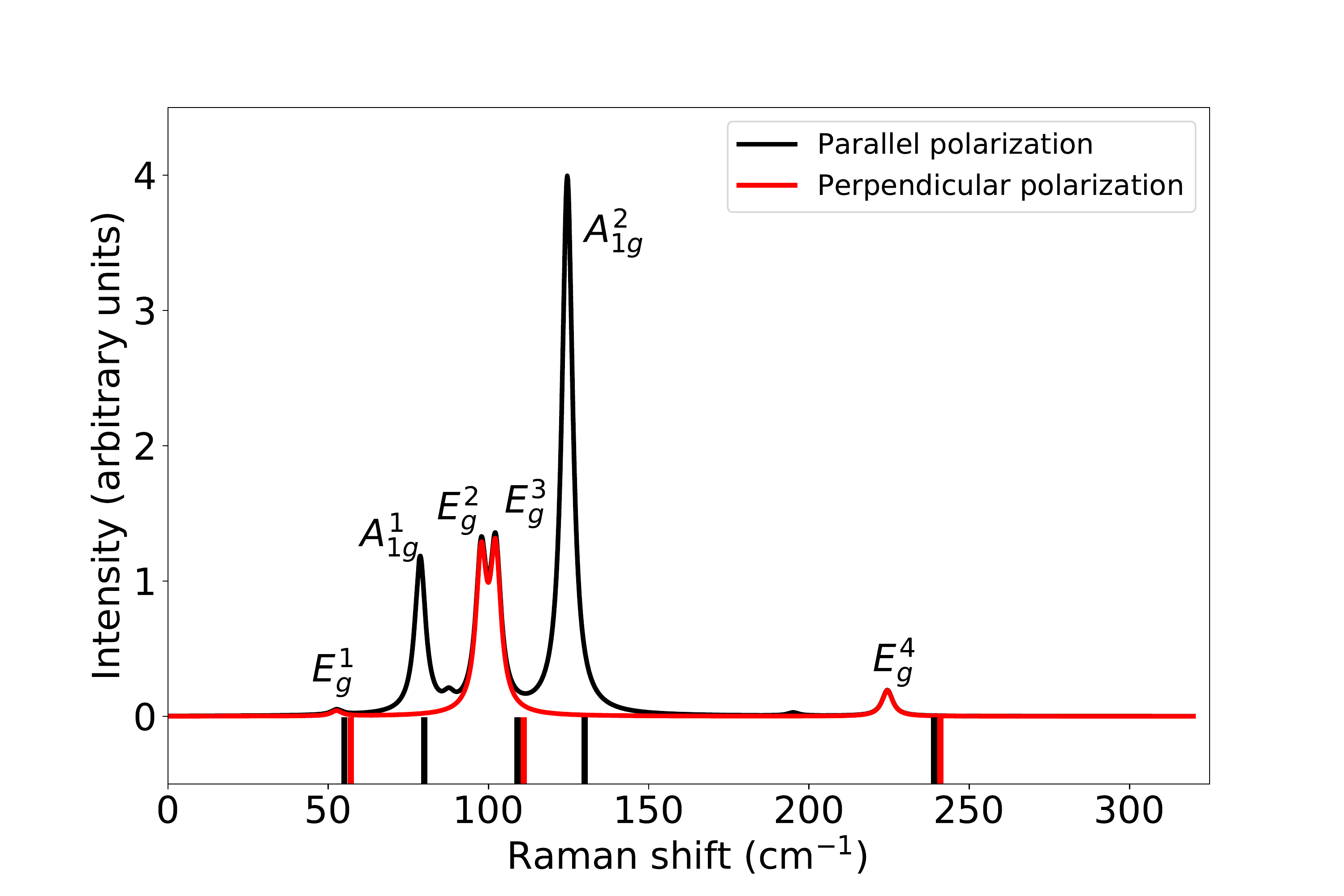}
\caption{(Color online) Raman spectrum for bulk R$\overline{3}$
  CrI$_3$ labeled with $D_{3d}$ irreducible representations for the
  active modes for parallel (black) and perpendicular (red)
  polarization. The peaks have been broadened by a Lorentzian with
  FWHM 4 cm$^{-1}$. The vertical lines beneath the spectrum show the
  locations of the major peaks in the measured Raman
  spectrum~\cite{seylerPC}.}
\label{fig:spectrum}
\end{figure}

We calculate the Raman tensor by taking a finite difference of the
macroscopic dielectric tensors for unit cells that have been displaced
in the positive and negative direction of each phonon mode. The
intensity of the Raman peaks with different polarizations are
proportional to the squared magnitude of various components of the
Raman tensor. Fig.~\ref{fig:spectrum} shows the calculated Raman
peaks for a bulk R$\overline{3}$ structure (low temperature) for both
parallel and perpendicular polarizations. The peaks have been labeled
with the appropriate $D_{3d}$ representations. The vertical lines
beneath the theoretical spectrum indicate the locations of the
experimentally measured Raman peaks from a sample of CrI$_3$ at $T=5$
K~\cite{seylerPC}. Comparison with measured peaks shows that all the
predicted peaks are present and that the $A_{1g}$ peaks vanish for
perpendicular polarization.

\begin{figure}
\includegraphics[width=\columnwidth]{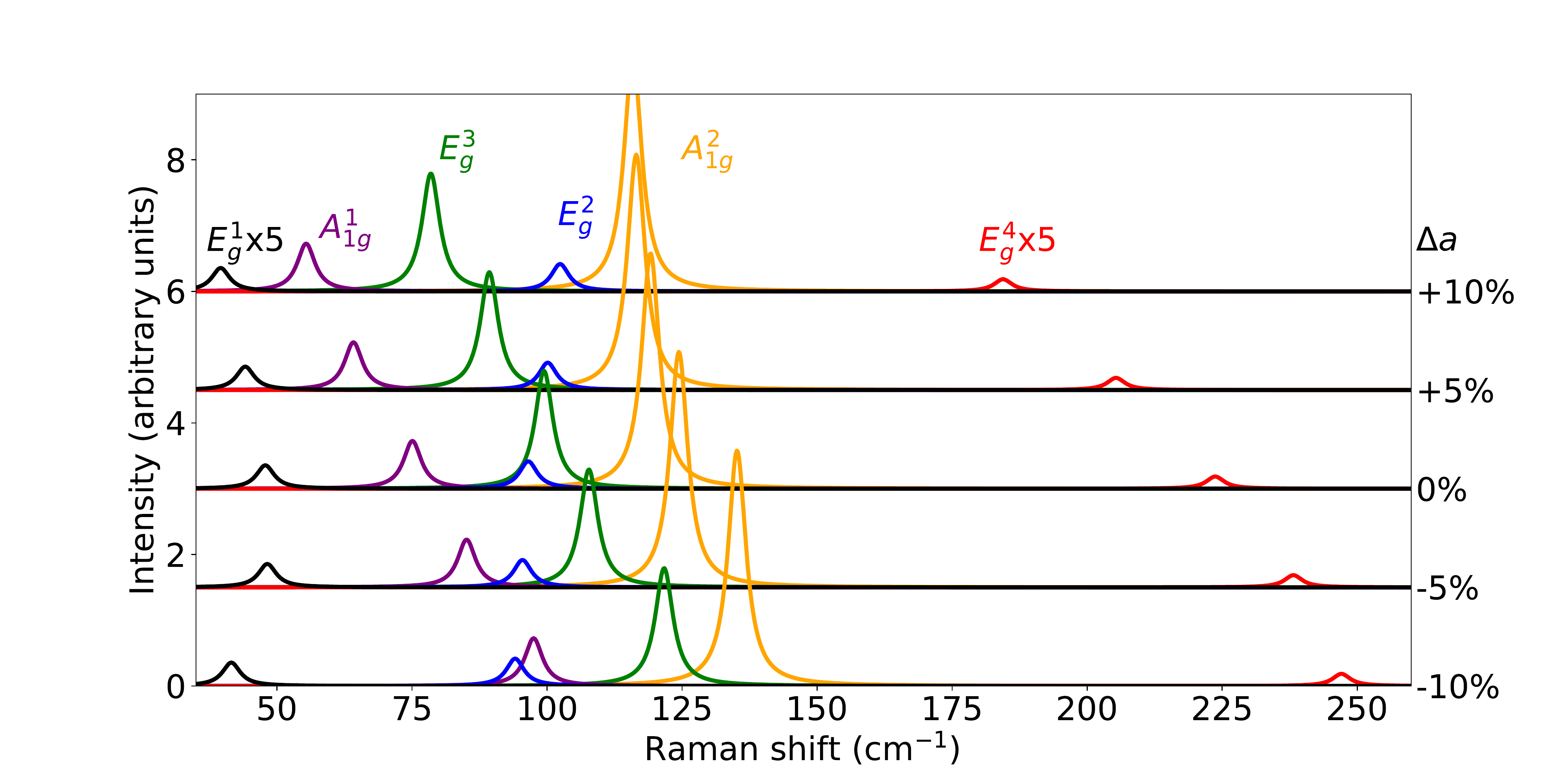}
\caption{(Color online) Change in single-layer R$\overline{3}$ CrI$_3$
  Raman peak locations due to uniform bi-axial strain. Peak positions
  are plotted for changes in the lattice constant ($\Delta a$) from
  $-10\%$ to $+10\%$ and offset in the vertical direction for
  clarity. The peaks have been broadened with a FWHM of 4 cm$^{-1}$.}
\label{fig:strain}
\end{figure}

When few- or single-layer materials are stacked or grown on
substrates, interactions between layers can cause strain on the atomic
structure, leading to changes in the phonon frequencies and observed
Raman spectra. To study the effects of strain on CrI$_3$ we calculate
the locations of the Raman peaks for a single layer in the
R$\overline{3}$ structure subject to uniform, bi-axial strain, induced
by increasing or decreasing the lattice constant by up to 10\%. The
changes in peak positions are shown in Fig,~\ref{fig:strain}, where
the Raman active modes are colored-coded and the spectra for different
lattice constants are offset in the vertical direction for
clarity. While $A_{1g}^1$, $A_{1g}^2$, $E_g^3$, and $E_g^4$ are all
blue shifted with increasing compression (decreasing lattice
constant), $E_g^2$ shows the opposite trend, and $E_g^1$ changes
non-monotonically. This suggests that the pattern of peaks and their
polarization dependence can be used as an indicator of the amount and
type of strain experienced by a sample \emph{in situ}.

\begin{figure}
  \includegraphics[width=\columnwidth]{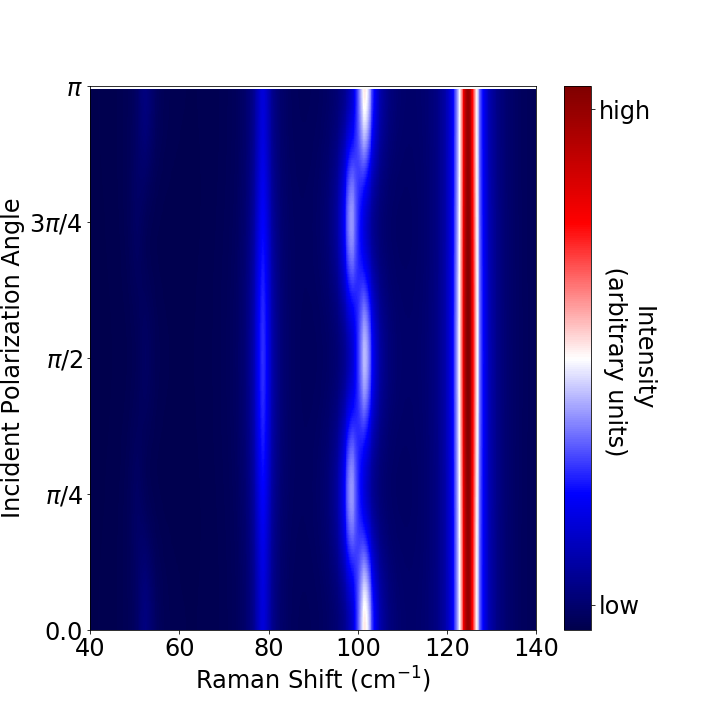}
  \caption{(Color online) Dependence of CrI$_3$ Raman intensity on incident
    angle for the bulk C2/m structure with parallel polarization. For
    perpendicular polarization the pattern is the same but the
    uniform peaks near 80 and 120 cm$^{-1}$ are absent.}
\label{fig:ang-dep}
\end{figure}

The Raman intensity is a function of the angle of the incident plane
polarization.  Our theoretical predictions for the angular dependence
in both R$\overline{3}$ and C2/m structures are the following:
\begin{itemize}
  \item[(i)] There is no discernible angular dependence for the R$\overline{3}$
structure. This occurs because the two degenerate modes within each
$E_g$ representation vary oppositely with angle but have the same
magnitude and therefore
compensate for each other, producing signals that are uniform in the
incident angle.
\item[(ii)] In the bulk C2/m structure the same modes are not
protected by the crystal symmetry and both $E_g^2$ and $E_g^3$
representations are split. Within each pair, the two modes are out of
phase and have very different magnitudes, which allows the angular
dependence of one mode of each pair to dominate. The dominant mode of
$E_g^2$ is also out of phase with the dominant mode of $E_g^3$, which
gives rise to the pattern shown in Fig.~\ref{fig:ang-dep}.
\end{itemize}

\begin{figure*}
  \includegraphics[width=\columnwidth]{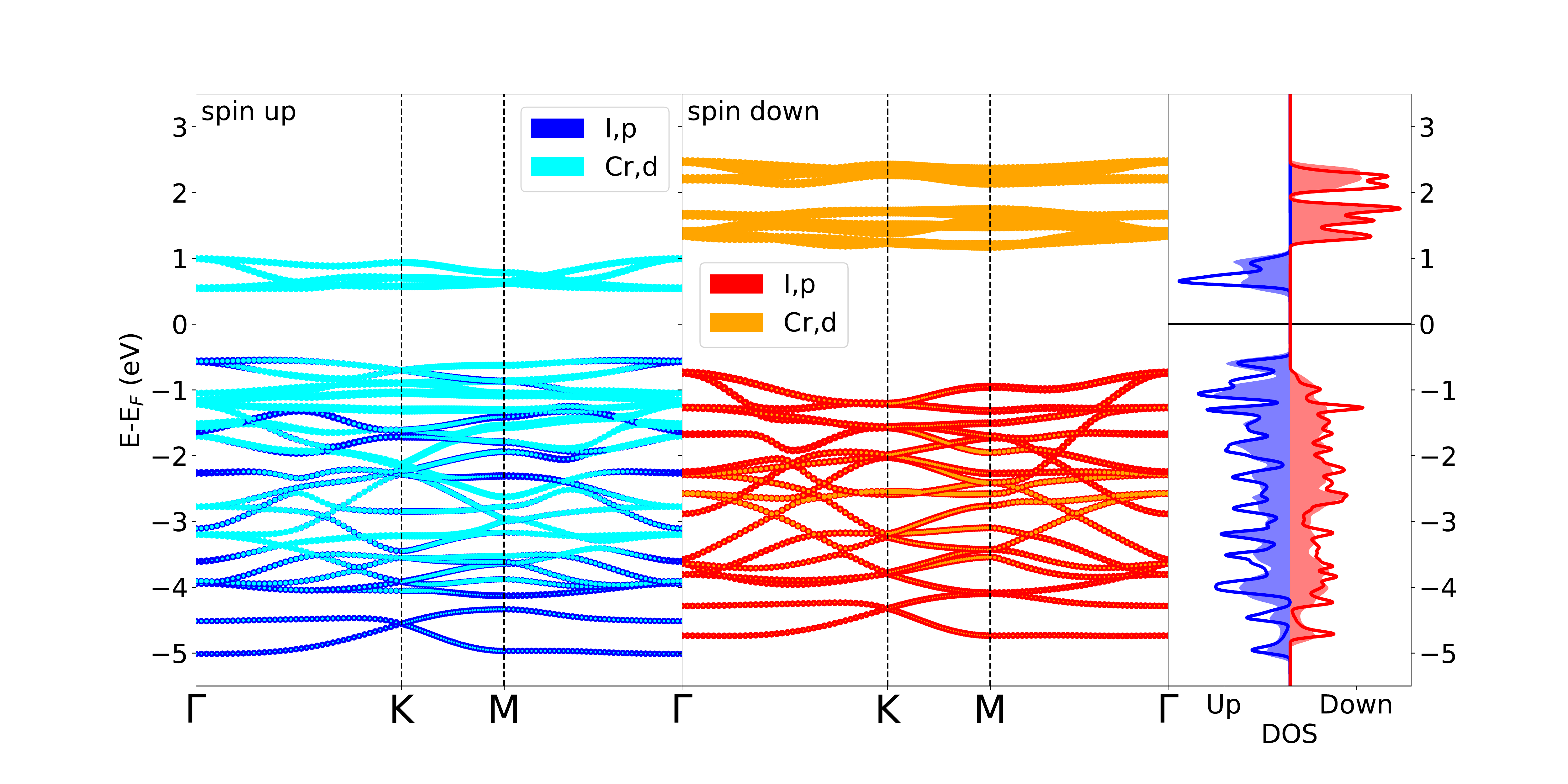}
\caption{(Color online) Orbital character of the electronic bands for
  the single-layer C2/m structure and DOS for the bulk (shaded
  regions) and single-layer (solid lines) structures.  }
\label{fig:c2m-elect}
\end{figure*}

Observation of non-trivial angular dependence in a sample at low
temperatures (R$\overline{3}$ structure) would be evidence of another
source of symmetry breaking, possibly time reversal symmetry breaking,
due to the ferromagnetic ordering of the spins. A similar effect has
been observed in Cr$_2$Ge$_2$Te$_6$~\cite{tian2016magneto} and used as
evidence for significant spin-phonon coupling. Indeed, the electronic
structure of a single layer shows that, for both the high- and
low-temperature structures the calculated magnetization is 3 $\mu_B$
per Cr ion and exhibits clear ferromagnetic ordering, as shown in
Fig.~\ref{fig:c2m-elect} for the single layer C2/m structure. The
qualitative features of the DOS and band structure are quite similar
for bulk and low-temperature structures.

In conclusion, our calculations of the Raman spectrum of CrI$_3$ show good
agreement with measured locations of
the Raman peaks and their polarization dependence. The different
amounts by which each mode shifts in response to strain suggests that
identifying the shifts in Raman peaks could be a straightforward
method for determining strain in experimental samples.  The dependence
on the incident polarization angle shows that measurements of the
angular dependence can be a sensitive probe of symmetry breaking
effects, which might be intimately tied to magnetic ordering.

\section*{Acknowledgments}

We thank E. Navarro-Moratalla for helpful discussions about
experimental studies of CrI$_3$ and K. Seyler and X. Xu for sharing with us
their Raman measurements. For the calculations we used
the Odyssey cluster supported by the FAS Division of Science, Research
Computing Group at Harvard University.

\bibliography{CrI3-raman}
\end{document}